\documentclass[aps,eqsecnum,nofootinbib,showpacs]{revtex4-1}
\usepackage[dvips]{graphicx}  
\usepackage{amsmath,amsfonts}
\usepackage{bm}
\newcommand{\romanNum}[1]{\@roman{#1}}
\usepackage{color}  
\def\be{\begin{eqnarray}}
\def\ee{\end{eqnarray}}
\def\beq{\begin{equation}}
\def\eeq{\end{equation}}

\def\p{\partial}

\def\({\left (}
\def\){\right )}
\def\[{\left [}
\def\[{\right ]}

\bmdefine{\bmk}{\bm{k}} \bmdefine{\bmx}{\bm{x}}
\bmdefine{\bmA}{\bm{A}} \bmdefine{\bmB}{\bm{B}}
\bmdefine{\bmJ}{\bm{J}}

\newcommand{\calO}{\mathcal{O}}

\newcommand{\pdiffII}[2]{ \frac{\partial^2 #1}{\partial #2^2} }
\begin{document}

\author{Hua-Bi Zeng$^{1}$, Zhe-Yong Fan$^{1}$, and Hong-Shi Zong$^{1,2}$}
\address{$^{1}$ Department of Physics, Nanjing University, Nanjing 210093, China}
\address{$^{2}$ Joint Center for Particle, Nuclear Physics and Cosmology, Nanjing 210093, China}
\title{$d$-wave Holographic Superconductor Vortex Lattice and Non-Abelian Holographic Superconductor Droplet}

\begin{abstract}
A $d$-wave holographic superconductor in the presence of a constant magnetic
field is studied by perturbation method. We obtain both droplet and triangular vortex lattice solutions. The results are the same as that of an $s$-wave holographic superconductor. The non-Abelian holographic superconductor with $p+ip$-wave background in the presence of a magnetic field is also studied. Unlike the $d$-wave and $s$-wave models, it is found that the non-Abelian model has only droplet solution.
\end{abstract}
\maketitle

\section{ introduction}
The correspondence between a $d$ dimensional quantum field theory and a $d+1$ dimensional gravity theory has provided a new method to understand the strong coupled field theory \cite{1,2,3,4}. The application of this duality to condensed matter physics is helpful to understanding the strong coupled many-body systems. Now there are many
attempts to use the Gauge/Gravity correspondence to study superfluidity/superconductivity  \cite{5,6,7,8,9,42,10,11,12,13,14,15,16,17,18} (see, for example, Refs. \cite{19,20,21,22} for reviews). The holographic superconductors are possible since there are classic gravity theories in AdS space which show local $U(1)$ symmetry broken solutions below a critical
temperature \cite{23,24}. Therefore, the dual field theories break the global $U(1)$ symmetry and they can be used to study superfluidity or superconductivity (in the limit that the $U(1)$ symmetry is gauged).

The initial holographic superconductor is an $s$-wave one because the order parameter is
a scalar. After that follow the non-Abelian holographic superconductors with vector
parameters which can be dual to a $p$-wave or a $p+ip$-wave superconductors.
The $s$-wave holographic model couples the Abelian Higgs model to gravity
with a negative cosmological constant. One can get solutions which spontaneously break the Abelian gauge symmetry via a charged complex scalar condensate near the horizon of the black hole when the temperature is low enough. The behavior of an $s$-wave holographic superconductor in the presence of a magnetic field has been studied in many papers
\cite{25,26,27,28,29,30,31,32}. The vortex solution for this model
has been constructed in Refs. \cite{28,29,30,31}. Especially, Maeda, Natsuume and Okamura
analytically obtain the same Abrikosov lattice solution as that in the Ginzburg-Landau theory  \cite{31}. These results indicate that this $s$-wave holographic superconductor
is of type II. The coherence length $\xi$ is studied in Ref. \cite{27}. $\xi$ shows the Ginzburg-Landau behavior $(1-T/T_c)^{-1/2}$ at the phase transition point.
Recently, a $d$-wave (spin two) holographic superconductor has been constructed, in which the complex scalar field in the $s$-wave model is replaced by a symmetric traceless tensor field whose condensate spontaneously breaks the gauge symmetry below $T_c$, and becomes zero and so that the symmetry is restored above $T_c$ \cite{18}.  We found that the critical exponents of the correlation function and the penetration length at $T_c$ take the
mean-field theory values \cite{33}. Another holographic superconductor model of $d$-wave gap was given in Refs. \cite{34,35,36}.

The action of the non-Abelian holographic superconductor model consists of $SU(2)$ gauge fields and the Einstein-Hilbert action. This Einstein-Yang-Mills (EYM) theory with fewer parameters whose Lagrangian is determined by symmetry
principles is constructed by Gubser \cite{24} and is
shown to have spontaneous symmetry breaking solutions due to a
condensate of non-Abelian gauge fields in the theory. Gubser and
Pufu studied this model with both $p$-wave backgrounds and
$(p+ip)$-wave backgrounds \cite{16}. Roberts and Hartnoll
studied the $(p+ip)$-wave backgrounds and found two major
nonconventional features for this holographic superconductor which
are different from their $s$-wave counterpart. One is the existence of
a pseudogap at zero temperature, and the other is the spontaneous
breaking of the time reversal symmetry \cite{17}. The zero temperature limit of the model is studied in Ref. \cite{37}, while in Refs. \cite{38,39} the model including back-reactions is discussed. In our recent paper \cite{40}, we studied the phase transition properties of
this model in the presence of a constant external magnetic field. We found that the
added background magnetic field indeed suppresses the superconductivity. Following closely
Maeda and Okamura \cite{27}, we studied the superconducting coherence length and magnetic penetration depth of the $p$-wave holographic superconductor by using perturbation theory near the critical temperature in Ref. \cite{41}. The results are the same as the case of the $s$-wave holographic superconductor which has been studied in Ref. \cite{27}.  The method that we used to study holographic superconductors is called the "bottom-up" approach: we put an arbitrary set of fields into the bulk AdS description and then study the bulk theory to get the information on the boundary field theory.
It is important to embed the holographic superconductors into string theory.
In Ref. \cite{43} the authors have studied the top-down approach considering various D-brane configurations in the AdS black hole
background in the string theory framework for the $p+ip$-wave and $s$-wave model. Scalars and gauge fields are common in all string
theory realizations of AdS/CFT and it is reasonable that $p+ip$-wave and $s$-wave holographic superconductors can be consistently embedded in a string
theory setting, whereas the embedding of the $d$-wave model with tensor field into string theory is still an open question \cite{36}.

In this paper, following the method used by Maeda, Natsuume and Okamura in Ref. \cite{31}, we analytically study the spatially dependent equations of motion for the $d$-wave and $p+ip$-wave holographic superconductor when the added magnetic field is slightly below the upper critical magnetic field. The following are our main results. Firstly, the upper critical magnetic field $B_{c2}$ for both two models is calculated and the phase diagrams are obtained.
Secondly, we get the same Abrikosov vortex lattice solutions for the $d$-wave model as that of the $s$-wave model. Thirdly, for the non-Abelian superconductor with $p+ip$ wave
backgrounds, we obtain the droplet solutions, but the vortex lattice solutions appearing in the $s$-wave and $d$-wave models are not possible here. The reason is that the Maxwell fields in the non-Abelian holographic superconductor are a subgroup of the $SU(2)$ gauge group,
hence they do not couple with the condensed fields via covariant derivative like the other two models.

The outline of the paper goes as follows. Section II is devoted to the construction of the triangle vortex solution of the $d$-wave model. In section III we discuss the $p+ip$-wave model's droplet solution. Finally, the conclusion and some discussion are given in Section IV.

\section{droplet solution and vortex lattice solution for the $d$-wave holographic superconductor}
In this section we first give the spatial dependent equations of motions for the $d$-wave model in the presence of an uniform magnetic field, then we construct the vortex lattice solution.

The full gravity theory in 3+1 dimensional spacetime which is dual to a 2+1 dimensional $d$-wave superconductor has the following action \cite{18}
\begin{eqnarray}
S &=&\frac{1}{2\kappa ^{2}}\int d^{4}x\sqrt{-g}\left\{ \left( R+\frac{6}{%
L^{2}}\right) +\mathcal{L}_{m}\right\} ,  \notag \\
\mathcal{L}_{m} &=&-\frac{L^{2}}{q^{2}}\left[ (D_{\mu }B_{\nu \gamma
})^{\ast }D^{\mu }B^{\nu \gamma }+m^{2}B_{\mu \nu }{}^{\ast }B^{\mu \nu }+%
\frac{1}{4}F_{\mu \nu }F^{\mu \nu }\right] ,  \label{Lagrangian}
\end{eqnarray}%
where $B_{\mu \nu}$ is a symmetric traceless tensor, $R$ is the Ricci scalar, the $6/L^{2}$ term gives a negative
cosmological constant and $L$ is the AdS radius. $\kappa ^{2}=8\pi G_{N}$ is the
gravitational coupling. $D_{\mu }$ is the covariant derivative in
the black hole background ($D_{\mu }=\partial _{\mu }+iA_{\mu }$ in
flat space), $q$ and $m^{2}$ are the charge and mass squared of
$B_{\mu \nu }$, respectively.

Working in the probe limit in which the matter fields do not back
react on the metric as in Ref. \cite{18} and taking the planar Schwarzchild-AdS  ansatz, the  black hole metric reads (we use mostly plus signature for the metric)
\begin{equation}
ds^2=-f(r)dt^2+\frac{dr^2}{f(r)}+\frac{r^2}{L^2}(dx^2+dy^2),
\label{metric}
\end{equation}
where the metric function $f(r)$ is
\begin{equation}
f(r)=\frac{r^2}{L^2}(1-\frac{r_0^3}{r^3}).
\end{equation}
$L$ and $r_0$ are the radius of the AdS spacetime and the horizon radius of the black hole, respectively. They determine the Hawking temperature of the black hole,
\begin{equation}
T=\frac{3r_0}{4\pi L^{2}},
\end{equation}
which is also the temperature of the dual gauge theory living on the
boundary of the AdS spacetime. Now we introduce a new coordinate $z=r_0/r$. The metric
(Eq. (\ref{metric})) then becomes
 \begin{equation}
ds^2=\frac{L^2\alpha^2(T)}{z^2}(-h(z)dt^2+dx^2+dy^2)+\frac{L^2dz^2}{z^2h(z)},
 \end{equation}
in which $h(z)=1-z^3$ and $\alpha(T)=r_0/L^2=4\pi T/3$.

For the $d$-wave backgrounds, the spatial dependent ansatz takes the following form \cite{27}
\begin{equation}
B_{\mu \nu }=\text{diagonal}\left( 0,0,f(z,x,y),-f(z,x,y)\right) ,~~A=\phi
(z,x,y)dt+A_y(z,x,y).
\end{equation}
We assume the vector potential $A_y$ is nonvanishing since we need a non-vanishing magnetic field on the boundary. $A_x$ can be set to zero when we take a suitable gauge.
This ansatz for the tensor field captures the feature of a $d$-wave superconductor in
which there is a condensate on the $x$-$y$ plane on the boundary
with translational invariance, and the rotational symmetry is broken
down to $Z(2)$ with the condensate changing its sign under a $\pi
/2$ rotation on the $x$-$y$ plane.

With this ansatz, we can derive the equations of motion:
\begin{equation}
h\partial_{z}^2 f+(\partial_{z}h
+\frac{2h}{z})\partial_{z}f+\frac{1}{\alpha^2}\partial_{x}^2 f+\frac{1}{\alpha^2}\partial_{y}^2 f+\frac{2iA_y}{\alpha^2}\partial_y f+\frac{if}{\alpha^2}\partial_y
A_y+\frac{2f\partial_z h}{z}+\frac{f \phi^2}{\alpha^2
h}-\frac{4fh}{z^2}-\frac{A_y^2f}{\alpha^2}-\frac{L^2m^2f}{z^2}=0,
\end{equation}
\begin{equation}
\alpha \partial_{z}^2 \phi+\frac{1}{\alpha
h}(\partial_x^2+\partial_y^2)\phi-\frac{4z^2 \mid f\mid ^2
\phi}{\alpha^3L^2h}=0,
\end{equation}
\begin{equation}
\alpha h\partial_z^2 A_y+\alpha \partial_z h\partial_z
A_y+\frac{1}{\alpha}\partial_x^2 A_y+\frac{2iz^2f^*\partial_y
f}{\alpha^3L^2}-\frac{2iz^2f\partial_y f^*}{\alpha^3L^2}-\frac{4z^2A_y \mid f\mid ^2}{\alpha^3L^2}=0.
\end{equation}
In order to solve the above equations, we have to introduce the following boundary conditions on the horizon and the boundary:

(i) On the horizon($z=1$), the scalar potential $\phi=0$ since the $\phi dt$ must be well defined. The other fields should be regular.

(ii) On the boundary ($z=0$), we are only interested in the $L^2m^2=-1/4$ case. The boundary conditions for $f$, $\phi$ and $A_y$ are \cite{27}:
\begin{align}
  & f= A_0 z^{\Delta_{-}}+A_1 z^{\Delta_{+}}
  + \cdots~,
\end{align}
\begin{align}
  & \phi = \mu - \rho z + \cdots~,
\end{align}
\begin{align}
  & B(\textbf{x})=\partial_x A_y - \partial_y A_x~,
\end{align}
in which  $\Delta_{\pm}=\frac{-1 \mp \sqrt{17+L^2m^2}}{2}$ and  $L^2m^2 \geq -4$. In the case we studied $\Delta_{-}=-5/2$ and $\Delta_{+}=3/2$.

$A_{0}$ is the source, then $A_{1}$ is the vacuum expectation
value (VEV) of the operator that couples to $B$ in the boundary theory.
$A_0$ can be set to zero \cite{18}. The order
parameter of the boundary theory can be read off from the asymptotic
behavior of tensor field $B_{\mu \nu}$,
\begin{equation}
\langle \mathcal{O}_{ij}\rangle =\left(
\begin{array}{cc}
A_{1} & 0 \\
0 & -A_{1}%
\end{array}%
\right)   \label{VEV}
\end{equation}%
where $(i,j)$ are the indexes in the boundary coordinates $(x,y)$.
$\mu$ is the chemical potential and $\rho$ is the charge density of the
field theory. $B(\textbf{x})$ is the magnetic field of the field theory on the boundary.

To exactly solve the above nonlinear coupled partial differential equations is a difficult task. But we can perturbatively solve these equations when the magnetic field
is slightly below the upper critical field $B_{c2}$. First we define a small parameter
$\epsilon=(B_{c2}-B)/B_{c2}$, then we can expand the fields as :
\begin{subequations}
\label{expansion}
\begin{align}
  & f({\bm x},z)
  = \epsilon^{1/2}f_1({\bm x},z) + \epsilon^{3/2}f_2({\bm x},z)
  + \cdots,
\\
  & A_y({\bm x},z)
  = A^{(0)}_y({\bm x},z) + \epsilon A^{(1)}_y({\bm x},z) + \cdots,
\\
&\phi({\bm x},z)=\phi^{(0)}({\bm x},z)+\epsilon\phi^{(1)}({\bm x},z)+ \cdots~
\end{align}
\end{subequations}
in which $\bm x=(x,y)$. The zeroth order solution corresponding to the normal state is
\begin{equation}
  f=0~~,\phi = \mu (1-z),~~
 A^{0}_y = B_{c2} x.
\label{sol-A-zerothd}
\end{equation}
We can see clearly that the magnetic field on the boundary is $B_{c2}$.
Substituting Eq. (\ref{sol-A-zerothd}) into the equations of motion, with the following masatz $f_1(\bmx,z)=e^{ipy}F(x,z;p)/L$ ($p$ is a constant), the equation of motion for $F$ is
\begin{align}
  & \left[h\partial_{z}^2 +(\partial_{z}h
+\frac{2h}{z})\partial_{z}+\frac{2\partial_z h}{z}+\frac{\mu^2(1-z)^2}{\alpha^2
h}-\frac{4h}{z^2}-\frac{L^2m^2f}{z^2}~\right] F(x,u;p)
\nonumber \\
  =&~\frac{1}{\alpha^2} \left[ - \pdiffII{}{x}
  + \left( p - B_{c2} x \right)^2~\right] F(x,u;p) .
\label{eq-psi-first}
\end{align}
Then we separate the $F$ as $F_n(x,z; p)=\rho_n(z)\gamma_n(x;p)/L$, where
$\lambda_n$ is a constant. $\rho_n$ and $\gamma_n$ admit the following equations:
\begin{subequations}
\begin{align}
  & \left( - \pdiffII{}{X} + \frac{X^2}{4} \right)
  \gamma_n(x;p)
  = \frac{\lambda_n}{2}\, \gamma_n(x;p) ,
\label{eq-gamma} \\
  &h\partial_{z}^2 +(\partial_{z}h
+\frac{2h}{z})\partial_{z} \rho_n(z)
\nonumber \\
  &\hspace*{0.5truecm}
  = \left( \frac{m^2 L^2}{z^2}
  - \frac{q^2}{h} (1-z)^2+\frac{4h}{z^2}-\frac{2\partial_z h}{z}
  + q^2 \frac{B_{c2} \lambda_n}{\mu^2}
  \right) \rho_n ,
\label{eq-rho}
\end{align}
\end{subequations}
where $X := \sqrt{2 B_{c2}} ( x - p/B_{c2} )$£¬ $q:=\mu/\alpha$
are dimensionless, $L^2m^2=-1/4$ in our calculation. Eq. (II.17a) determines the distribution of the order parameter on the
$x-y$ plane, while Eq. (II.17b) determines when a superconducting phase transition will
occur.

The solution of (\ref{eq-gamma}) that satisfies the boundary condition and $\lim_{|x|\to\infty
}|\gamma_n|<\infty$ can be expressed in terms of the Hermite functions $H_n$ as follows
\begin{align}
\gamma_n(x;p) = e^{- X^2/4 } H_n(X),
\end{align}
and the corresponding eigenvalue $\lambda_n$ is
\begin{align}
\lambda_n = 2 n + 1 ,
\end{align}
where $n=0,1,2,3 \cdots$. The $n=0$ solution is the droplet solution, and the vortex solution can be constructed from the droplet solution:
\begin{align}
  & \gamma_0(x;p)
  = e^{- X^2/4}
  = \exp\left[ - \frac{1}{2 r_0^2}
    \left(x - p r_0^2 \right)^2 \right],
\label{sol-droplet}
\end{align}
where $r_0 := 1/\sqrt{B_{c2}}$.

Before we construct the vortex lattice solution, let us discuss the phase diagram.
From Eq. (II.17b) we can obtain the phase diagram. The upper critical magnetic field given by this equation has a non-zero solution satisfying the boundary conditions. This can be done numerically for a given $q$, which corresponds to
a fixed temperature. We can find a critical value $B_{c2}/\mu^2$ above which the equation have only vanishing solution. The maximum upper critical magnetic field is given for $n=0$ when ($\lambda_n$) take the minimum value. In Fig. 1 we give the phase diagram, from which we can find that $B_{c2} \propto (1-B/B_{c2})$ around $T_c$. This is the same as the BCS theory.
\begin{figure}[!htbp] \centering
\includegraphics[width=3.0in]{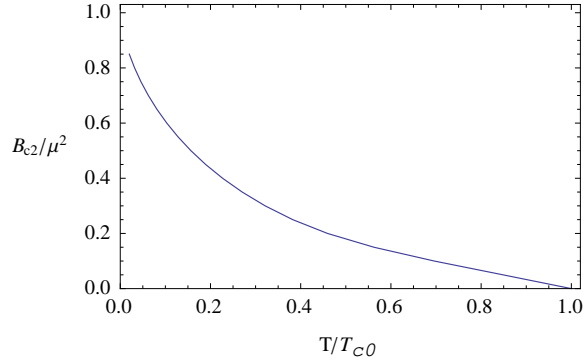}
\hspace{2.2cm}\caption{The phase diagram of a $d$-wave holographic superconductor in the presence of a magnetic field.}
\end{figure}

Since $\lambda_n$ is independent of $p$,
a linear superposition of the solutions $e^{ipy}\rho_0(u)\gamma_0(x;p)$
with different $p$ is also a solution of the equation of motion
for $f_1$:
\begin{equation}
 f_1({\bm x},u)
  = \frac{\rho_0(u)}{L}
  \sum_{l} c_l\, e^{i p_l y} \gamma_0(x; p_l) .
\label{eq:lattice_sol} \\
\end{equation}
Here we get the most important result in this section. When we choose a suitable configuration of $c_l$ and $p_l$, we can construct triangular lattice solutions. It is very interesting that the result Eq. (II.21) is very similar to the
expression of the order parameter of G-L theory for the type II superconductor in the presence of a magnetic field when $B=B_{c2}$, which is
\begin{equation}
\psi_L=\sum_{l} c_l e^{i p_l y}\textrm{ exp}[-\frac{x-x_l}{2 \xi^2}],
\end{equation}
where $\xi$ is the superconducting coherence length, $x_l=\frac{k \Phi_0}{2\pi B}$, and $\Phi_0$ is the flux quantum. Comparing Eq. (II.21) with Eq. (II.22), we get
\begin{equation}
B_{c2}\propto \frac{1}{\xi^2},
\end{equation}
which is also similar to the result of the GL theory. According to the behavior that $B_{c2}\propto(1-T/T_c)$ near $T_c$, we have $\xi \propto (1-T/T_c)^{-1/2}$. This result is also the same as that of the GL theory. We have also obtained this result by another way in Ref. \cite{33}.

Thus, the construction of triangular lattice from droplet solutions is similar to what Abrikosov did in his initial paper. This procedure has been made for the $s$-wave model in Ref. \cite{31}. In the $d$-wave model, the construction process is the same. We briefly review the result below, considering the following form of $p_l$ and $c_l$:
\begin{subequations}
\begin{align}
  & f_1({\bm x},u)
  = \frac{\rho_0(u)}{L}
  \sum_{l=-\infty}^{\infty} c_l\, e^{i p_l y} \gamma_0(x; p_l) ,
\label{eq:lattice_sol} \\
  & c_l
  := \exp\left( - i \frac{\pi}{^2} l^2 \right),
  \hspace{1.0truecm}
   p_l := \frac{2 \pi l}{a_1 r_0},
\label{eq:def-C_l-p_l}
\end{align}
\label{superpose-psi}%
\end{subequations}

\noindent
for arbitrary parameters $a_1$. The solution in Eq. (\ref{superpose-psi})
represents a lattice.
$\sigma(\bm{x}) := |\gamma_L(\bm{x})|^2$
in which the fundamental region $V_0$ is spanned by two vectors
${\bm b}_1=a_1 r_0\p_y$ and
${\bm b}_2=2\pi r_0/a_1\p_x+a_1 r_0/2\p_y$,
and the area is given by $2\pi r_0^2$.
Then the magnetic flux penetrating the unit cell is given by
$B_{c2} \times (\text{Area}) = 2 \pi$.
This shows the quantization of the magnetic flux penetrating a vortex.

The order parameter vanishes at
\begin{align}
  & \bm{x}_{m,n} = \left( m + \frac{1}{2} \right) \bm{b}_1
  + \left( n + \frac{1}{2} \right) \bm{b}_2 ,
\end{align}
for any integers $m$, $n$. The phase of $\langle \mathcal{O} \rangle \propto \gamma_L(\bmx)$ rotates by $2\pi$ around each $\bm{x}_{m,n}$~. When
\be
\label{triangle-parameters}
\frac{a_1}{2}=3^{-1/4}\sqrt{\pi},
\ee
the three adjoining vortices ${\bm x}_{m,n}$
form an equilateral triangle, which is the triangular vortex lattice solution.
Here, following \cite{18}, in our calculation we only focus on the case of $L^2m^2=-1/4$.
Since different allowed values of mass only change the dimension of the condensation
operator in Eq.(II.10), it will not affect the Eq. (II.17a) and also our construction of
vortex solution. So we can conclude that for other values of $m$, the vortex solutions
are also expected.

\section{$p+ip$-wave holographic superconductor droplet solution}
First we review the gravity dual theory of the non-Abelian holographic
superconductor with $p+ip$-wave background. The full EYM theory in 3+1
dimensional spacetime considered in Refs. \cite{16,17} has the following action
\begin{eqnarray}
S_{\textmd{EYM}}=
\int\sqrt{-g}d^4x\left[\frac{1}{2\kappa_4^2}\left(R+\frac{6}{L^2}\right)
-\frac{L^2}{2g_{\rm YM}^2}\textmd{Tr}(F_{\mu\nu}F^{\mu\nu})\right],
\end{eqnarray}
where $g_{\rm YM}$ is the gauge coupling constant and
$F_{\mu\nu}=T^aF^a_{\mu\nu}=\partial_\mu A_\nu-\partial_\nu
A_\mu-i[A_\mu,A_\nu]$ is the field strength of the gauge field
$A=A_\mu dx^\mu=T^aA^a_\mu dx^\mu$. For the $SU(2)$ case,
$[T^a,T^b]=i\epsilon^{abc}T^c$ and
$\textmd{Tr}(T^aT^b)=\delta^{ab}/2$, where $\epsilon^{abc}$ is the
totally antisymmetric tensor with $\epsilon^{123}=1$. The Yang-Mills
Lagrangian becomes $\textmd{Tr}(F_{\mu\nu}F^{\mu\nu})=F^a_{\mu\nu}F^{a\mu\nu}/2$ with
the field strength component $F^a_{\mu\nu}=\partial_\mu
A^a_\nu-\partial_\nu A^a_\mu+\epsilon^{abc}A^b_\mu A^c_\nu$.

Working in the probe limit in which the matter fields do not
backreact on the metric as in Refs. \cite{16,17}
and taking the planar Schwarzchild-AdS  ansatz, the black hole
metric is the same as Eq. (II.2), in which $f(r)$ is given in Eq. (II.3).
The Hawking temperature of black hole is $T=\frac{3r_0}{4 \pi r_0}$.

Now we introduce a new coordinate $z=r_0/r$. The metric (Eq. (\ref{metric})) then becomes
 \begin{equation}
ds^2=\frac{L^2\beta^2(T)}{z^2}(-h(z)dt^2+dx^2+dy^2)+\frac{L^2dz^2}{z^2h(z)},
 \end{equation}
where $h(z)=1-z^3$ and $\beta(T)=r_0/L^2=4\pi T/3$.

Using the Euler-Lagrange equations, one can obtain the equations of
motion for the gauge fields,
\begin{equation}
\frac{1}{\sqrt{-g}}\partial_{\mu}\left(\sqrt{-g}F^{a\mu\nu}\right)
+\epsilon^{abc}A^{b}_{u}F^{c\mu\nu}=0.
\end{equation}
For the $p+ip$-wave backgrounds without a external
magnetic field, the ansatz \cite{16,17} takes the
following form,
\begin{equation}
A=\phi(z,x,y)T^3dt++w(z,x,y)T^1dx+w(z,x,y)T^2dy,
\end{equation}
in which we have included the spatial dependence.
Here the $U(1)$ subgroup of $SU(2)$ generated by $T^3$ is identified
with the electromagnetic gauge group \cite{16,17} and $\phi$ is
the electrostatic potential. Thus the black hole can carry charge through the
condensate $w$, which spontaneously breaks the $U(1)$ gauge
symmetry below a critical temperature. This is a Higgs mechanism, but there are Goldstone bosons corresponding to changing the directions of the condensate in real
space or gauge space. They must be visible in the bulk as normal
modes or (more likely) quasi-normal modes.

In order to add a homogenous magnetic field on the boundary (where the
field theory lives), we also need non-vanishing $A_x^3(z,x,y)$ and $A_y^{3}(z,x,y)$.
Together with these non-vanishing terms above, the equations of motions for $w,\phi,A_y^3,A_x^3$ are:
\begin{equation}
2\alpha\partial_z(h\partial_z w)+\frac{1}{\alpha}(\partial_x^2 +\partial_y^2 )w+\frac{2}{\alpha h}\phi^2w-\frac{2}{\alpha}w^3
-\frac{3}{\alpha}w\partial_x A_y^3+\frac{3}{\alpha}w\partial_y A_x^3-\frac{2}{\alpha}w((A_x^3)^2+(A_y^3)^2)=0
\end{equation}
\begin{equation}
-\alpha\partial_z^2\phi-\frac{1}{\alpha h}(\partial_x^2\phi+\partial_y^2\phi)+\frac{2}{\alpha h}\phi w^2=0
\end{equation}
\begin{equation}
\frac{1}{\alpha}\partial_x^2 A_y^3+\alpha\partial_z(h\partial_z A_y^3)-\frac{1}{\alpha}\partial_x(\partial_y A_x^3)
+\frac{3}{\alpha}w\partial_x w-\frac{1}{\alpha}w^2 A_y^3=0
\end{equation}
\begin{equation}
\frac{1}{\alpha}\partial_y^2 A_x^3+\alpha\partial_z(h\partial_z A_x^3)-\frac{1}{\alpha}\partial_y(\partial_x A_y^3)
-\frac{3}{\alpha}w\partial_x w-\frac{1}{\alpha}w^2 A_x^3=0
\end{equation}
the boundary conditions are:

(i) For $w$, it should be regular at the horizon. At the boundary,
the asymptotic behavior of $w$ has the following expression
\begin{align}
  & w = \frac{\langle
\calO\rangle}{\sqrt{2}}z
  + \cdots~.
\end{align}
where $\langle \calO\rangle$ is the condensate  of the charged
operator dual to the field $w$ and is the order parameter for the
superconductivity phase. Here we demand the constant term
vanish in Eq. (III.9) since we require that there be no source term in the
field theory action for the operator $\langle \calO \rangle$
\cite{16,17}. In fact, it is a requirement for the absence of
such a term which in principle can be present.

(ii) For the electromagnetic gauge fields $\phi$, $A_x^3$ and $A_y^3$, at the boundary we have
\begin{align}
  & \phi = \mu/\beta(T) - q z + \cdots~, B(\textbf{x})=\partial_x A_y^3 - \partial_y A_x^3,
\end{align}
in which $\mu$ is the chemical potential and $q$ is the charge density, while $ B(\textbf{x})$ is the magnetic field. Obviously, at the horizon, we need $\phi=0$, and $A_x^3$ and $A_y^3$ are both regular. Now, our task is to solve this equation to get the information we need.

Just as above, to exactly solve these non-linear coupled differential equations is also very difficult. However, as we did in the last section, we can solve the equations analytically by perturbation method near the upper critical magnetic field $B_{c2}$. Above $B_{c2}$
there is no condensation at any temperature. As in the last section, we define a deviation parameter $\epsilon=(B_{c2}-B)/B_{c2}$. When $B$ is slightly below the upper critical magnetic field, we can expand the four fields $w$, $\phi$, $A_x^3$ and $A_y^3$ as:
\begin{equation}
w=\epsilon^{1/2}w_1+\epsilon^{3/2}w_2 + \cdots~
\end{equation}
\begin{equation}
\phi=\phi^{(0)}+\epsilon\phi^{(1)}+ \cdots~
\end{equation}
\begin{equation}
A_y^3=A_y^{3(0)}+\epsilon A_y^{3(1)}+ \cdots~
\end{equation}
\begin{equation}
A_x^3=A_x^{3(0)}+\epsilon A_x^{3(1)}+ \cdots~
\end{equation}
Note that all these fields are functions of $x,y,z$.
The zeroth order solutions which correspond to the normal states ($w=0$) with fixed chemical
potential and magnetic field $B_{c2}$ are
\begin{equation}
\phi=\mu(1-z), A_y^{3(0)}=B_{c2}x, A_x^{3(0)}=0.
\end{equation}
Substituting the expansion into the equation for $w$ and making a separation of variables, we have $w(x,y,z)=e^{ipy}m(x,z;p)$. For a constant $p$, we get the equation of motion for $m(x,z;p)$
\begin{equation}
[2\alpha^2 \partial_z(h\partial_z)+\frac{2\mu^2(1-z)^2}{h}]m(x,z;p)=[-\partial_x^2+p^2+B_{c2}^2 x^2+3B_{c2}]m(x,z;p)
\end{equation}
We separate the variable $m$ as $m_n(x,z;p)=\rho_n(z)\gamma_n(x;p)$ with a separation constant $\lambda_n$. The equations of motion for $\rho_n(z)$ and $\gamma_n$ are:
\begin{equation}
   \left( - \pdiffII{}{X} + \frac{X^2}{4} +\frac{p^2}{2B_{c2}}+\frac{3}{2}\right)
  \gamma_n(x;p)
  = \frac{\lambda_n}{2}\, \gamma_n(x;p) ,
\end{equation}
and
\begin{equation}
\partial_z(h\partial_z \rho_n(z))=\left (-\frac{q^2}{h}(1-z)^2+q^2\frac{B_{c2}\lambda_n}{2\mu^2}\right)\rho_n(z),
\end{equation}
where $X := \sqrt{2 B_{c2}} x$ and $q:=\mu/\alpha$
are dimensionless. Eq. (III.17) gives the spatial profile while Eq. (III.18) gives the upper critical magnetic field. It is easy to see that the regular and bounded solution of
Eq. (III.17) can be expressed in terms of the Hermite function $H_n$:
\begin{align}
  & \gamma_n(x;p) = e^{- X^2/4} H_n(X),
\end{align}
and the corresponding eigenvalue $\lambda_n$ is
\begin{align}
  & \lambda_n = 2 n + 4+\frac{p^2}{B_{c2}},
\end{align}
for a non-negative integer $n$.

We can see that the solution of Eq. (III.19) is independent of $p$, which is different
from the $s$-wave model in Eq. (II.18). For the $s$-wave one, the spatially dependent solutions $\gamma_n(x;p)$ are functions of $p$, and the vortex lattice solutions with a periodicity in $x$ direction can be constructed by a superposition
of different solutions for different $p$ when $n=0$. This difference leads us to conclude that the non-Abelian holographic superconductors cannot have vortex lattice solutions.

These solutions are actually droplet solutions in the sense that they fall off rapidly at large $|x|$. A single droplet solution can be obtained by considering another zeroth order solution rather than Eq. (III.15). We consider the following zeroth order solution:
\begin{equation}
\phi=\mu(1-z), A_y^{3(0)}=B_{c2}x/2, A_x^{3(0)}=-B_{c2}y/2,
\end{equation}
which satisfies the equations of motion.
With this solution, after a separation of variables $w(x,y,z)=\gamma_n(x)\gamma_m(y)\rho_{m,n}(z)$, the solutions for
the three fields are
\begin{equation}
   \left( - \pdiffII{}{X} + \frac{X^2}{8} +\frac{3}{2}\right)
  \gamma_m(x)
  = \frac{\lambda_m}{2}\, \gamma_m(x) ,
\end{equation}
\begin{equation}
   \left( - \pdiffII{}{Y} + \frac{Y^2}{8} +\frac{3}{2}\right)
  \gamma_n(y)
  = \frac{\lambda_n}{2}\, \gamma_n(x) ,
\end{equation}
\begin{equation}
\partial_z(h\partial_z \rho_{m,n}(z))=\left (-\frac{q^2}{h}(1-z)^2+q^2\frac{B_{c2}(\lambda_n+\lambda_m)}{2\mu^2}\right )\rho_{m,n}(z)
\end{equation}
where $X := \sqrt{2 B_{c2}} x,Y := \sqrt{2 B_{c2}} y $ and $q:=\mu/\alpha$
are dimensionless.
The solution for $\gamma_m(x)$ and $\gamma_n(x)$ are
\begin{align}
  & \gamma_n(x;p) = e^{- X^2/8} H_n(X),
\end{align}
\begin{align}
  & \gamma_m(y;p) = e^{- Y^2/8} H_m(X),
\end{align}
and the corresponding eigenvalue $\lambda_n$ is
\begin{align}
  & \lambda_m = 2 m + 4,
\end{align}
\begin{align}
  & \lambda_n = 2 n + 4.
\end{align}
The order parameter for the field theory is given by the boundary value of $\partial_z w=\partial_z\rho_{m,n}\gamma_n(x)\gamma_m(y)$
for $z=0$. $\rho_{m,n}$ is given by Eq. (III.18), which is independent of $x$ and $y$, and according to this equation the upper critical magnetic field has
a non-zero solution satisfying the boundary conditions. This can be done numerically as we did in the last section, for a given $q$, which corresponds to
a fixed temperature. We can also find a critical value $B_{c2}/\mu^2$ above which the equation has only the vanishing solution. The maximum upper critical magnetic field is obtained when ($\lambda_n+\lambda_m$) takes the minimum value ($m=n=0$).
The single droplet solution is also obtained when $m=n=0$, which is $\gamma_n(x)\gamma_m(y)= e^{-(Y^2/8+ X^2/8)}$. In Fig. 2 we give the phase diagram.
\begin{figure}[!htbp] \centering
\includegraphics[width=3.0in]{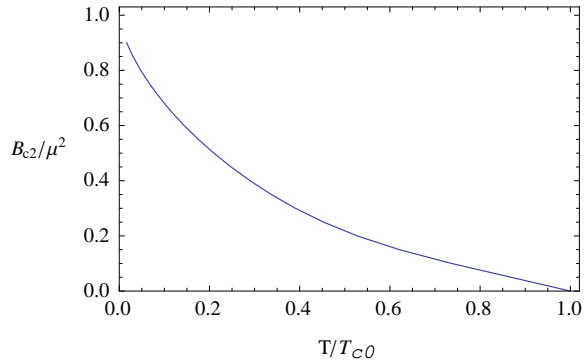}
\hspace{2.2cm}\caption{Phase diagram of the $p+ip$-wave holographic superconductor in the presence of a magnetic field, in which we also find $B_{c2}\propto(1-T/T_c)$ at $T_c$, just like the BCS theory. It can also be seen that the phase diagram is very similar to the $d$-wave one.}
\end{figure}
\section{ conclusion and discussion}
A $d$-wave and a $p+ip$ wave holographic superconductors are studied by
an analytic perturbation method around the upper critical magnetic field.
The $d$-wave model has the same droplet and triangular vortex lattice solutions as the $s$-wave one, and the lattice solution is constructed by the superposition of droplet solutions.
The $p+ip$-wave model has only droplet solutions, because the $x$ direction property is independent of $p$ (see Eq. (III.19)) and so a superposition of droplet solutions will not give the lattice solution. According to details of our calculation, this is due to the fact that in the non-Abelian model the Maxwell fields appear as a $U(1)$ subgroup of the $SU(2)$ field and they do not couple with the condensed charged fields via the covariant derivative as in the $s$-wave and $d$-wave models. Technically, the difference
between the two situations can be traced to the nature of the minimal
coupling of the charged condensates to the gauge potential.
The HTSC is believed to be $d$-wave pairing and it is indeed a type II superconductor
with vortex solutions in the presence of a magnetic field as the
holographic $d$-wave one. The real $p+ip$-wave superconductor will also enter into the vortex state when a magnetic field is present, which is clearly different from the holographic $p+ip$ model. When we interpret this $p+ip$-wave model to be a superfluidity, the computations described here can equally apply to this case as well with a slight modification. In this case, the rotation of the superfluid is analogous to the magnetic field. The experimental realization of superfluid droplets was given in Ref. \cite{44}.
Although the holographic superconductor models have many similar behaviors as the real ones, it is still an open question to bridge the gap between the holographic models and the field-theoretic method for superconductors.
Finally, since a magnetic field can penetrate both holographic superconductors, both of them should be a type II superconductor.
\section{acknowledgment}
We would thank Wei-min Sun for valuable comments. This work is supported in part by the National Natural Science
Foundation of China (under Grant Nos. 10775069, 10935001 and 11075075) and the Research Fund
for the Doctoral Program of Higher Education (Grant No. 20080284020).

\end{document}